\documentclass{ws-procs9x6}

\def\F{\mathcal{F}}
\def\K{\mathcal{K}}
\def\Q{$Q^2$}
\def\G{GeV$^2$}
\begin{document}

\title{Baryon Form Factors at High Momentum Transfer and
GPD's}

\author{PAUL STOLER}

\address{Physics Department, Rensselaer Polytechnic Institute, Troy NY12180}

\maketitle

\abstracts{Nucleon elastic and transition form factors at
high momentum transfer $-t$ are treated in terms of generalized parton
distributions in a two-body framework.
In this framework the high $-t$ dependence of the form factors
give information about the high   $k_\perp$ , or short distance
$b_\perp$ correlations of nucleon model wave functions. Applications 
are made to elastic and resonance nucleon form factors, and
real Compton Scattering.}

    During the past several years there has been considerable
discussion of how to describe exclusive reactions at momentum
transfers which are experimentally attainable. 
While pQCD is an interesting mechanism which probes the
simplest Fock state component of the hadron, most theoretical
studies agree that even at the highest attainable momentum
transfers, there is a large {\em soft} contribution which involves
more complex components of the hadronic wave functions.
The so-called handbag~\cite{efremov} mechanism has 
evolved to describe such
soft processes, and achieves its full power at high momentum transfer
where a process can be factorized into a fully perturbative
hard amplitude  and a 
{\em generalized parton distribution}
 (GPD)~\cite{ji}~\cite{rad_gpd}~\cite{collins}, which 
represents the off-diagonal probability of the  interacting quark 
being placed back into
the remaining hadron, keeping it in-tact at a different transfered
longitudinal momentum. The power of the mechanism is that the same
soft GPD, which contains the information about the hadronic structure
is accessed in a variety of different reactions, while the hard 
perturbative part is reaction specific. The GPD's give us unique
information about the longitudinal ($x$) and transverse ($k_\perp$)
parton momentum distributions, and importantly, about the
interference between the initial parton wave function and the phase
shifted final parton wave function.

The GPD approach manifests itself in two kinematical regimes,
corresponding to the $t$ dependent {\em form factor} type reaction,
and the $t \to t_{min}$ {\em off-forward} production of mesons or photons.
Here we focus on the former.
In such a reaction the incident real or virtual 
photon interacts perturbatively with one of the quarks within
the hadron, which 
is re-absorbed into the hadron leaving it in-tact or in a higher
resonant state.  This is a Feynman type  reaction
which involves the full complexity of the non-perturbative nucleon
structure, as opposed to the leading order pQCD mechanism, which involves
only the valence quark Fock state.
 Form factors are the $x$ moments of
the GPD's, and as such 
constrain the longitudinal dependence of the nucleon 
structure. As a function of $t$ they  uniquely 
constrain the $k_\perp$ dependence of the
nucleon's wave functions.  Fourier transforms of the GPD's - 
$\F_b(x,\vec b_\perp) \propto  \int \vec{dq}_\perp 
exp(i \vec b_\perp\cdot \vec q_\perp) \F(x,t)$ -,
directly give the transverse spatial impact parameter distribution 
of the quarks for each longitudinal momentum fraction~\cite{burkardt}.
Thus, together with $x$ distributions obtained in DIS
the  $k_\perp$ accessed in form factor measurements give us
a unique 3 dimensional picture of the quark distributions
in the nucleon.
Examples of 
 reactions accessible via GPDs include  the nucleon elastic
Dirac and Pauli form factors $F_1$ and $F_2$
(or equivalently  $G_{Ep}$ and $G_{Mp}$), resonance transition
amplitudes such as $A_{1/2}$ for $N\to S_{11}(1535)$, or $G_M^*$
for  $N\to \Delta$, and 
Compton scattering form factors
$R_V$ and $R_A$  and their polarization  asymmetries.
The relationship
of the GPD's to these various form factors is given as
follows:

For elastic scattering 
 
\begin{equation}
F_1(t)=\int^1_0\sum_q \F^q(x,t)dx  
\hspace{0.3in}
F_2(t)=\int^1_0\sum_q \K^q(x,t)dx.
\label{eq:F12}
\end{equation}

\noindent where $q$ signifies both quark and anti-quark flavors.
We work in a reference frame in which the total
momentum transfer is transverse so that  $\zeta$=0, and denote
$ \F^q(x,t) \equiv \F^q_0(x,t)$, 
\ $ \K^q(x,t) \equiv \K^q_0(x,t)$.

 For Compton scattering~\cite{rad_wacs}

\begin{equation}
R_1(t)=\int^1_0\sum_q{1\over x} \F^q(x,t)dx 
\hspace{0.3in}
R_2(t)=\int^1_0\sum_q{1\over x} \K^q(x,t)dx.
\label{eq:R12}
\end{equation}

Resonance transition form factors access components of the
GPD's which are not accessed in elastic scattering or Compton
scattering. The  $N\to\Delta$ form factors are related
to isovector components of the GPD's~\cite{frankfurt}~\cite{polyakov}. 

\begin{equation}
 G^*_M = \int^1_0\sum_q \F^q_M(x,t)dx\ \ \  
 G^*_E = \int^1_0\sum_q \F^q_E(x,t)dx\ \ \  
 G^*_C = \int^1_0\sum_q \F^q_C(x,t)dx 
\label{eq:delta}
\end{equation}

\noindent where  $G^*_M$, $G^*_E$ and  $G^*_C$ are magnetic, electric
and Coulomb transition form factors~\cite{jones}, and 
$ \F^q_M$,  $\F^q_E$, and $\F^q_C$ are axial (isovector) GPD's,
which can be related to elastic GPD's in the large $N_C$ limit
through isospin rotations~\cite{polyakov}. The $N \to S_{11}$ transition 
form factor is also important, as it probes
fundamental aspects of dynamical chiral symmetry breaking in QCD.
If chiral symmetry were not broken, the $S_{11}$ would be the nucleon's
parity partner and the $N$ and $S_{11}$ masses would be degenerate.

\begin{figure}[h]
\centerline{\epsfxsize=3.5in \epsfbox{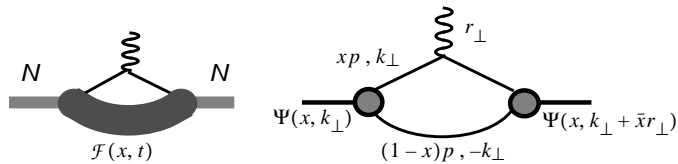}}   
\caption{Schematic relation between the two-body and handbag
mechanisms discussed in the text. \label{twobody}}
\end{figure}

As a basis for constructing the GPD's we use the two-body
model introduced in~\cite{rad_wacs} whose connection with
the handbag is illustrated in fig.~\ref{twobody}.

In this framework the GPD is written

\begin{equation}
\F(x,t) = \int {\Psi^{*}(x,k_\perp + \bar x r_\perp)\Psi(x,k_\perp)
{{d^2k_\perp}\over{16\pi^3}}}
\label{eq:Ftb}
\end{equation}

\noindent where $\bar x \equiv 1-x$,  

An example of a specific model wave function~\cite{stoler}
is 

\begin{equation} 
\Psi(x,k_\perp) = \Phi(x)\left( A_s e^{-k_\perp^2/2x\bar x \lambda^2}
+ A_h {{x\bar x \Lambda^2}\over{k_\perp^2 + \Lambda^2}} \right)
 \equiv \Psi_{soft}+\Psi_{hard}\\
\label{eq:psihard}
\end{equation}

\begin{figure}[ht]
\begin{minipage}[b]{1.5in}
\raggedright
\caption{\label{wf} 
The function $\Psi(k_\perp) \equiv \int{\Psi(x, k_\perp) dx}$ 
vs. $k_\perp$. The dashed curve
is due to the  soft Gaussian component $\Psi_{soft}$, with
$\lambda^2 = 0.6\ {\rm GeV^2}$. The solid curve is 
$\Psi_{soft} + \Psi_{hard}$, with $A_h$ = 0.24, 
$k_{\perp ,max}$ = 4 GeV, and
cutoff parameter $\Lambda$ = 0.45 GeV.
\vspace{0.4in}}
\end{minipage}
\hspace{0.3in}
\begin{minipage}[b]{2.0in}
\psfig{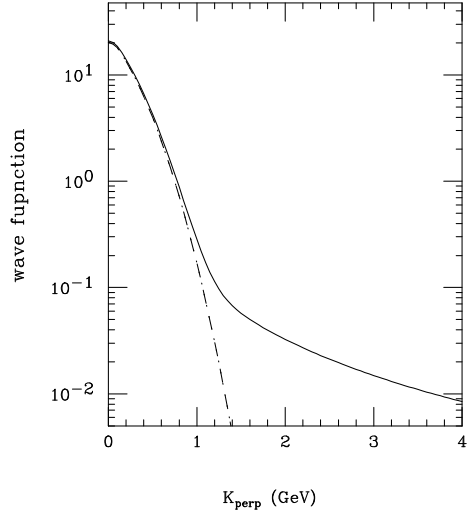}
\end{minipage}
\end{figure}

The function $\Phi(x)$ is constrained so that $\F(x,0)$ reduces to
the valence quark distribution $f(x)$. It was shown 
in ref.~\cite{stoler} that although a  Gaussian form of
the $k_\perp$ dependence in $\Psi_{soft}$
accounts for the magnitude and shape of the elastic  $F_1$ for
\Q\ below several \G, it is inadequate at higher \Q. 
However, the addition of a small  $\Psi_{hard}$
component in eq.~(\ref{eq:psihard}) can dramatically improve the
agreement at high \Q.
As an example of a power law dependence,
 we choose an {\em ad-hoc} $1/k_\perp^2$  behavior with
lower cutoff parameter $\Lambda$. A similar parameterization
is chosen for $F_2$ with $\K^q(x,0)=\sqrt{(1-x)}\F^q(x,0)$. In order to
constrain the parameters of eq.~(\ref{eq:psihard}) the available
data on both $G_{Mp}$ and $G_{Ep}/G_{Mp}$ were
simultaneously reproduced,
giving  $A_s = \sqrt{1 - A_h^2} =  0.97$, $A_H = 0.24$, 
$\lambda_1^2=0.6$ GeV$^2$  and  $\lambda_2^2=0.45$  GeV$^2$.
The function $\Psi(k_\perp)= \int{\Psi(k_\perp,x)dx}$ is shown in 
fig.~\ref{wf}. Only at $k_\perp$ greater than about 1 GeV does the
hard tail important.

\begin{figure}[ht]
\begin{minipage}[b]{1.25in}
\caption{ 
\label{GMp-tot-qb} Upper: Proton magnetic form factor 
$G_{Mp}/G_D$, where $G_D=1/(1+Q^2/0.71)^2$. Data are from
SLAC$^{9,10}$ with low energy data 
reevaluated$^{11}$. The dashed curve uses only
$\Psi_{soft}$, while the solid curve
uses $\Psi_{soft}+\Psi_{hard}$. Lower: The impact parameter 
dependence of the curves in the upper figure,
$G_{Mp}(b_\perp)= \int dx \F_b(x,b_\perp)$. The curve at the 
bottom left labelled ``hard tail'' is the difference between 
the solid and dashed curves, which is responsible for most of
the form factor at high \Q.
\vspace{0.4in}
}
\end{minipage}
\hspace{0.1in}
\begin{minipage}[b]{2.0in}
\psfig{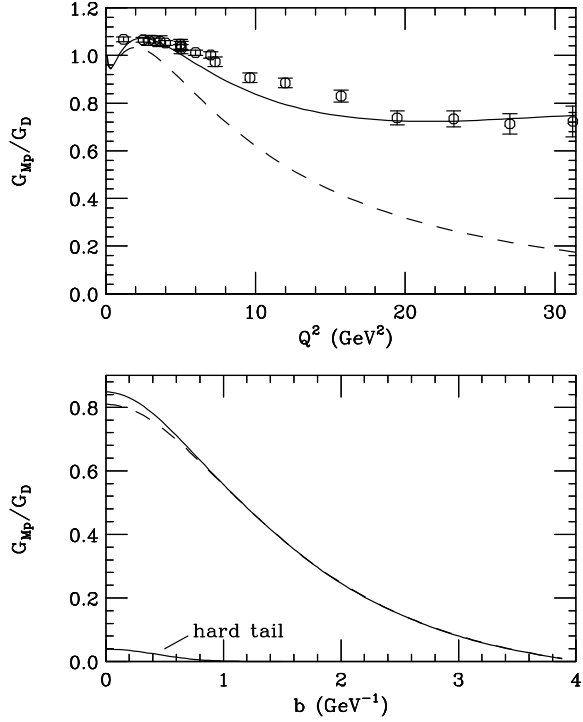}
\end{minipage}
\end{figure}

\begin{figure}[ht]
\begin{minipage}[b]{1.25in}
\caption{\label{gegmp} $GE_p/GM_p$ 
for $\Psi_{soft}+\Psi_{hard}$ and $\Psi_{hard}$ alone
are compared with the recent JLab 
data$^{12}$. The curves are as in fig.~3.
\vspace{0.5in}
}
\end{minipage}
\begin{minipage}[b]{2.0in}
\psfig{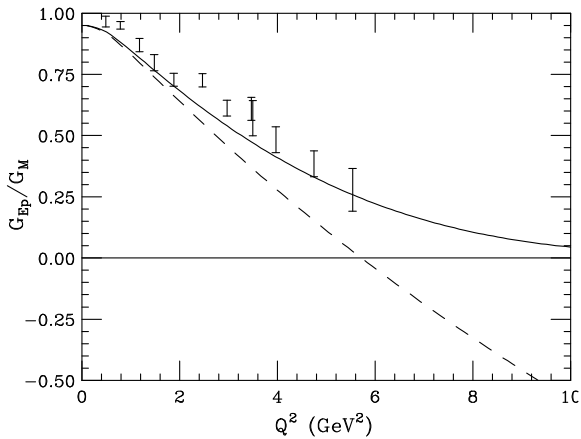}
\end{minipage}
\end{figure}

\begin{figure}[h]
\begin{minipage}[b]{1.25in}
\caption{ \label{GMDELTA} The $N\to\Delta$ magnetic form factor
$G_M^*(Q^2)$ relative to the dipole  $G_D=3/(1+Q^2/.71)^2$ }
\vspace{0.4 in}
\end{minipage}
\begin{minipage}[b]{2.0in}
\psfig{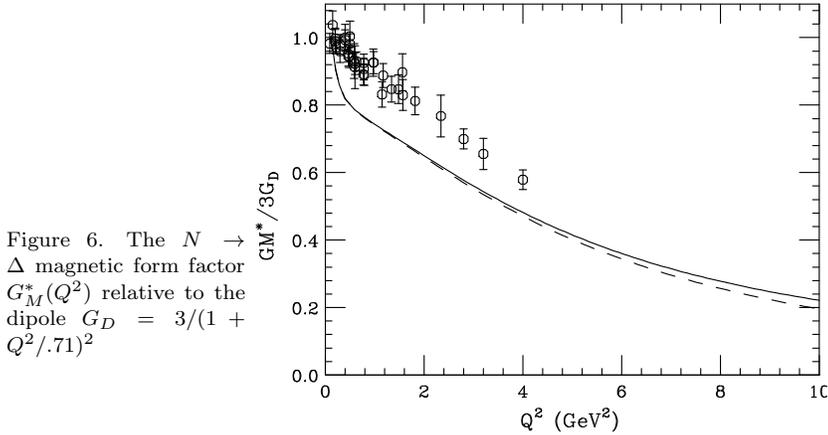}
\end{minipage}
\end{figure}

The fits to the data using respectively $\Psi=\Psi_{soft}+\Psi_{hard}$, 
and $\Psi=\Psi_{soft}$ are shown in 
figs.~\ref{GMp-tot-qb}~\ref{gegmp}~\ref{GMDELTA}.

As seen in the top panel of fig.~\ref{GMp-tot-qb},
this rather small addition of 
high momentum components can account for the high, as well
as the low \Q\ magnetic form factor. Interestingly,
ref.~\cite{carlson}found that even in a pQCD calculation  a power
law tail is useful in reproducing the high \Q\ data.

Taking the Fourier transforms of the GPD's gives the spatial impact
parameter distribution of the struck quarks. The bottom panel
in fig.~\ref{GMp-tot-qb} shows 
 
$$ \F_b(x,b_\perp) = \int dq_\perp e^{i \vec b_\perp\cdot \vec q_\perp} \F(x,t).$$

\noindent and the effect of $\Psi_{hard}$. Only a small addition of 
small impact parameter components to the wave function
accounts for most of the form factor at high \Q.

In fig.~\ref{gegmp} the obtained values of $GE_p/GM_p$ 
for $\Psi_{soft}+\Psi_{hard}$ and $\Psi_{hard}$ alone
are compared with the recent JLab data~\cite{perdrisat}.

\begin{figure}[h]
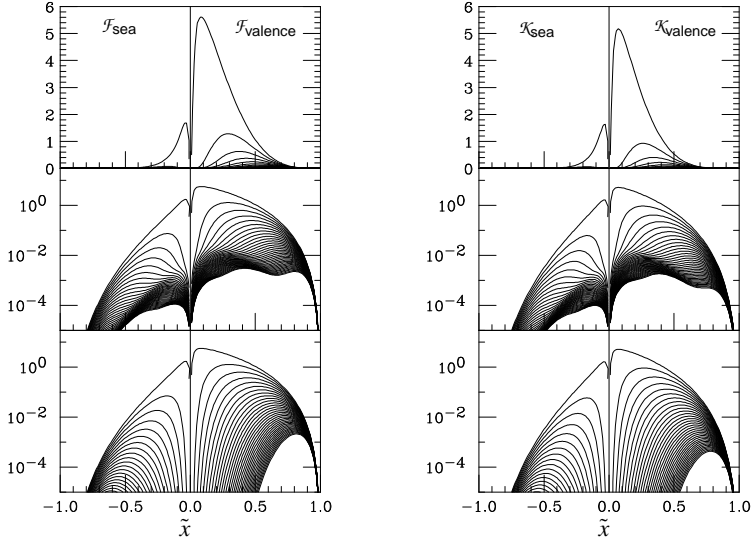

\begin{minipage}[b]{2.5in}
\psfig{file=F.epsi,width=1.7in}
\end{minipage}
\begin{minipage}[b]{2.5in}
\hspace{0.4in}
\psfig{file=K.epsi,width=1.7in}
\end{minipage}
\caption{
\label{gpd} GPD's as a function of $\tilde x$ for various
values of $t$, where $\tilde x = x$ (x-tilde) for valence quarks,
and  $\tilde x =  - x$ for the sea quarks. 
 The figures on the left and right 
are for  $\F$ and $\K$ respectively. The graphs
for  positive $\tilde x$ represent the {\em valence} quark contribution,
while the graphs for negative $\tilde x$ represent the {\em sea} quark
contributions.  The
individual curves range from $|t|$  $\sim 0$ GeV$^2$ (highest curve
in each panel)   
to  $|t|$  = 35  GeV$^2$ (lowest curve in each panel). 
The upper and middle panels
are the GPD's for the full wave function $\Psi_{soft}+\Psi{hard}$,
while those in the lowest panels are obtained using the
 $\Psi_{soft}$ soft only. Note that the addition
of the $\Psi_{hard}$ mainly affects the GPD's at higher $|t|$
and $\tilde x < 0.5$}
\end{figure}

The obtained GPD's as a function of $x$ and $t$ are shown in 
fig.~\ref{gpd}.

One may apply the constraints of the elastic form factors to
investigate properties of inelastic resonance transitions.
For example, in the large $N_c$ limit the GPDs for the $N\to \Delta(1232)$
transition are expected  to be  isovector components of the 
elastic GPD, which is approximately given by

$$\F^{(IV)}_M = {2\over{\sqrt{3}}}\K^{(IV)}_M={2\over{\sqrt{3}}}
\left(\K^u-\K^d\right),$$

\noindent where $\K^u$ and $\K^d$ are the GPD's for the up and down
quarks respectively.
Figure~\ref{GMDELTA} shows the result of applying the GPD's
from elastic scattering to the $N\to\Delta$ transition.
The data was renormalized by the ratio 3/2.14, to bring into
line the nucleon isovector form factor at \Q=0 with the experimental
value for the $N\to \Delta$.

In summary, it is seen that complete knowledge of the
various types of baryon form factors provides very
strong constraints for model wave functions and GPD's.

\end{document}